\documentclass[a4paper,fleqn,usenatbib]{mnras}
\usepackage{newtxtext,newtxmath}

\usepackage[T1]{fontenc}
\usepackage{ae,aecompl}

\usepackage{graphicx}	
\usepackage{amsmath}	
\usepackage{amssymb}	

\title[Single-pulse observations of PSR J1745$-$2900]{Single-pulse
  observations of the Galactic Center magnetar PSR J1745$-$2900 at 3.1
  GHz}

\author[W. M. Yan et al.]{
W. M. Yan,$^{1,2,3}$\thanks{E-mail: yanwm@xao.ac.cn (WMY)}
N. Wang,$^{1,2,3}$
R. N. Manchester,$^{4}$
Z. G. Wen$^{1,2,3}$
and J. P. Yuan$^{1,2,3}$
\\
$^{1}$Xinjiang Astronomical Observatory, CAS, 150 Science 1-Street, Urumqi, 
Xinjiang, 830011, China\\
$^{2}$Key Laboratory of Radio Astronomy, Chinese Academy of Sciences, 
Nanjing 210008, China\\
$^{3}$Xinjiang Key Laboratory of Radio Astrophysics, 150 Science 1-Street, Urumqi, 
Xinjiang, 830011, China\\
$^{4}$CSIRO Astronomy and Space Science, Australia Telescope National 
Facility, PO Box 76, Epping, NSW 1710, Australia
}

\date{Accepted XXX. Received YYY; in original form ZZZ}

\pubyear{2017}

\begin{document}
\label{firstpage}
\pagerange{\pageref{firstpage}--\pageref{lastpage}}
\maketitle

\begin{abstract}
We report on single-pulse observations of the Galactic Center magnetar
PSR J1745$-$2900 that were made using the Parkes 64-m radio telescope
with a central frequency of 3.1 GHz at five observing epochs between
2013 July and August. The shape of the integrated pulse profiles was
relatively stable across the five observations, indicating that the
pulsar was in a stable state between MJDs 56475 and 56514.  This
extends the known stable state of this pulsar to 6.8 months.  Short
term pulse shape variations were also detected. It is shown that this
pulsar switches between two emission modes frequently and that
the typical duration of each mode is about ten minutes.  No
giant pulses or subpulse drifting were observed. Apparent
  nulls in the pulse emission were detected on MJD 56500. Although
  there are many differences between the radio emission of 
  magnetars and normal radio pulsars, they also share some
  properties. The detection of mode changing and pulse nulling in
  PSR J1745$-$2900 suggests that the basic radio emission process for
  magnetars and normal pulsars is the same.
\end{abstract}




\section{Introduction} \label{sec:intro}

Magnetars are commonly considered to be rotating neutron stars whose
inferred surface dipolar magnetic fields are extremely strong
(typically $10^{14}$ -- $10^{15}$ G) \citep{dt92a}. Their X-ray
and $\gamma$-ray luminosities are usually orders of magnitude larger
than their spin-down luminosity. It is believed that magnetar
emission, particularly at high energies, is powered by the decay of the
enormous magnetic fields \citep{td95,td96a} instead of by the
spin-down. However, the discoveries of low magnetic field magnetars
\citep{ret+10,rie+12,rvi+14} may challenge the original assumption
that a high surface dipolar magnetic field strength is required for
the activity of a typical magnetar.

The magnetar group is classified by observers as Soft Gamma Repeaters
(SGRs) and Anomalous X-ray Pulsars (AXPs), and both subgroups are
typically detected at high energies.  The total number of magnetars
currently known is
29\footnote{\url{http://www.physics.mcgill.ca/~pulsar/magnetar/main.html}}
\citep{ok14}. To date, only four of them have shown pulsed radio
emmision \citep{crh+06,crhr07,lbb+10,sj13, efk+13}. PSR J1745$-$2900
is the newest radio-emitting magnetar. It was serendipitously
discovered by the {\tt\string Swift} telescope as an X-ray flare that
came from the region near Sagittarius A* (Sgr A*) \citep{kbk+13}.
Subsequent observations of PSR J1745$-$2900 by the {\tt\string NuSTAR}
telescope revealed pulsed X-ray emission with a spin period of 3.76~s
and a spin-down rate of $\dot{P} = 6.5 \times 10^{-12}$, which implies
a surface dipolar magnetic field $B = 1.6 \times
10^{14}\ \mathrm{G}$. This confirmed PSR J1745$-$2900 as a magnetar in
the Galactic center (GC) region \citep{mgz+13}.  With a series of
observations with the Chandra and the Swift telescopes, this pulsar
was later localized only $2\farcs4$ away from Sgr A* \citep{rep+13}.

Radio pulsations from PSR J1745$-$2900 were subsequently detected with
many radio tesescopes.  \citet{sj13} and \citet{efk+13} reported
multifrequency radio observations of PSR J1745$-$2900 and showed that
it has the largest dispersion measure (DM = $1778\ \pm
3\ \mathrm{cm^{-3}\ pc}$) and rotation measure (RM =
$-66960\ \pm\ 50\ \mathrm{rad\ m^{-2}}$) of any known pulsar. These
measurements constrain the strength of the magnetic field near the GC.
Based on high-resolution astrometry measurements for PSR J1745$-$2900
with the VLBA and VLA, \citet{bdd+14} found that the angular
broadening for this pulsar is in good agreement with that of Sgr A*,
confirming that PSR J1745$-$2900 and Sgr A* must be close to each
other to share a similar scattering medium.  The proper motion of PSR
J1745$-$2900 relative to Sgr A* was later measured by the VLBA
observations. \citet{bdd+15} demonstrated that this pulsar has a
transverse velocity of 236 km s$^{-1}$ at a projected separation of
0.097 pc from Sgr A*. Radio observations indicated that PSR
J1745$-$2900 has a relatively flat radio spectrum making the pulsar
detectable at millimeter bands \citep{tek+15,tde+17}.  Single-pulse
radio observations for PSR J1745$-$2900 were carried out by
\citet{laks15} and \citet{ysw+15} at frequencies above 8 GHz using the
GBT and the Shanghai Tian Ma Radio Telescope (TMRT)
respectively. Their results showed that the radio radiative activity
of the pulsar underwent a change from a fairly stable state to a more
erratic state. Both the flux density and the pulse profile morphology
showed substantial changes from epoch to epoch in the erratic
phase. No giant pulses or subpulse drifting were detected in these
observations.

The NE2001 model predicts a very large scattering timescale for PSR
J1745$-$2900. PSR J1745$-$2900 would be undetectable at frequencies
below 5 GHz if the prediction of the NE2001 model were true.  But the
observed scattering broadening was much lower than this prediction
\citep{sle+14,ppe+15}, implying that radio pulsations may be
detectable at relatively low frequencies. The new YWM16 model
\citep{ymw17} predicts the scattering observed in J1745$-$2900 very
well, based on the scattering results of \citet{kmn+15}.  In this
paper, we present the results of single-pulse observations at 3.1 GHz
for PSR J1745$-$2900 that were made with the Parkes 64-m radio
telescope, which is the lowest frequency single-pulse analysis for
this pulsar to date.  Details of the observing system and the
observations are given in Section~\ref{sec:obs}. The mean pulse
profile and single-pulse properties are shown in
Section~\ref{sec:results}. The implications of the results are
discussed in Section~\ref{sec:discn}.

\section{Observations} \label{sec:obs}

\begin{table*}
\begin{center}
\caption{Summary of the observations. Note that the symbols 
$\tau_{\mathrm{samp}}$ and T$_{\mathrm{obs}}$ represent the sampling 
interval and the duration of the observation respectively. For 
details about Project ID, see descriptions in the Parkes Pulsar 
Data Archive.}\label{tab:obs}
\begin{tabular}{cccccccc}
\hline
\hline
Date  & MJD &Project & Frequency & Bandwidth & No. of & $\tau_{\mathrm{samp}}$ & T$_{\mathrm{obs}}$ \\
(yyyy-mm-dd) & (d) &ID  & (MHz)     & (MHz)     & Channels & ($\mu$s)  & (min) \\
\hline
2013-07-02 & 56475 &P626  & 3100 & 1024 & 512 & 256 & 101 \\
2013-07-19 & 56492 &P574  & 3094 & 1024 & 512 & 256 & 20 \\
2013-07-21 & 56494 &P574  & 3094 & 1024 & 512 & 256 & 20 \\
2013-07-27 & 56500 &P626  & 3100 & 1024 & 512 & 128 & 160 \\
2013-08-10 & 56514 &P456  & 3094 & 1024 & 512 & 128 & 13 \\
\hline
\end{tabular}
\end{center}
\end{table*}

As the only known pulsar that is located at the GC, PSR J1745$-$1900
has been observed many times at multiple frequency bands by the Parkes
64-m radio telescope for many projects. Many of those data are
publicly available in the Parkes Pulsar Data
Archive\footnote{\url{https://data.csiro.au}} \citep{hmm+11}.  High
signal-to-noise ratio (S/N) and long duration are the essential
criteria for the selection of observational data to study single
pulses.  Under these criteria, five single-pulse observations of PSR
J1745$-$2900 made between 2013 July and August were found in the
Parkes Pulsar Data Archive and then analyzed in this paper.
Unfortunately, no suitable calibration observation can be found in the
Parkes Pulsar Data Archive for the five observations, so flux and
polarization calibration cannot be performed here.  The five
observations were taken with the 10-cm receiver, which has a bandwidth
of 1024 MHz centred around 3.1 GHz, and the fourth generation Parkes
digital filterbank system PDFB4.  Details of the observations are
summarized in Table~\ref{tab:obs}. The full bandwidth was divided into
512 channels to allow incoherent de-dispersion, resulting in a 2-MHz
channel width and hence a 0.98-ms dispersive time delay across each
frequency channel at 3.1 GHz for PSR J1745$-$2900 \citep{lk05}.  This
dispersion smear time is two orders of magnitude less than the
observed average half-power width of single pulses, so 512 channels
are sufficient for us.

The data were reduced using the {\tt\string DSPSR} package
\citep{vb11} to de-disperse and produce single-pulse integrations
which preserve information on individual pulses. The pulsar's
rotational ephemeris was taken from \citep{laks15} and the
single-pulse integrations were recorded using the {\tt\string PSRFITS}
data format \citep{hvm04} with 1024 phase bins per rotation
period. Radio frequency interference (RFI) in the single-pulse
archives was removed in affected frequency channels and time
sub-integrations using {\tt\string PAZ} and {\tt\string PAZI} programs
of the pulsar analysis system {\tt\string PSRCHIVE} \citep{hvm04}.
Even after this, the observed profile baselines varied significantly
probably because of residual low-level RFI. We removed the effect of
the varying baseline by subtracting the mean level in the off-pulse
window adjacent to the pulse from each single-pulse integration.  The
single-pulse analysis was carried out with the {\tt\string PSRSALSA}
package \citep{wel16} which is freely available
online\footnote{\url{https://github.com/weltevrede/psrsalsa}}.  The
single-pulse integrations were rebinned from 1024 to 512 pulse phase
bins to increase the S/N.

\section{Results} \label{sec:results}

We present and discuss the mean pulse profile and single-pulse
properties for PSR J1745$-$1900 in this section. As mentioned in
Section~\ref{sec:obs}, we cannot perform analyses of polarization
properties and absolute flux densities for this pulsar since no
calibration data are available.

\subsection{Mean pulse profiles} \label{subsec:mean}

\begin{figure*}
\centering
\includegraphics[angle=0,width=0.7\textwidth]{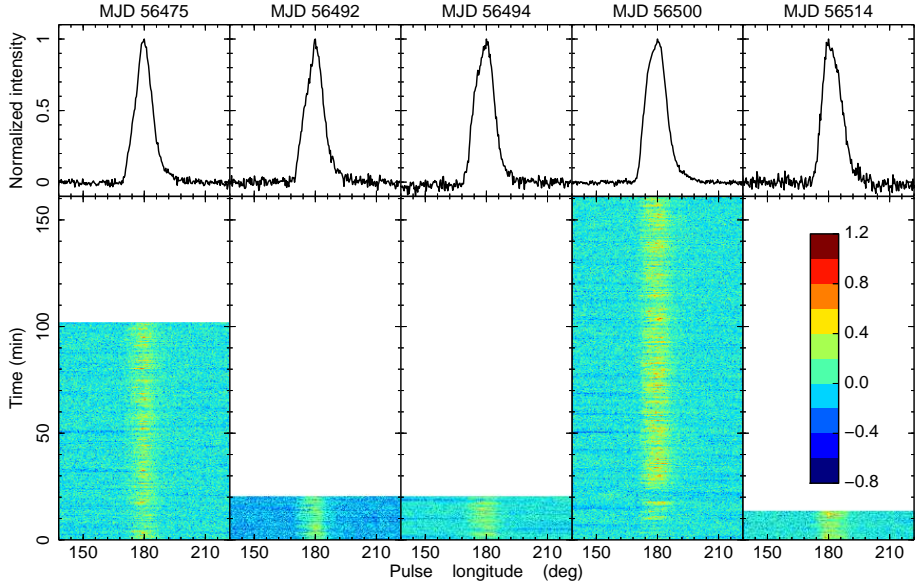}
\caption{3.1-GHz single-pulse stacks (lower) and corresponding mean pulse 
         profiles (upper) of PSR J1745$-$2900 observed at five observing epochs. 
         The flux density of each mean 
         pulse profile is normalized by its peak flux density. 
         The MJD for the day of observation is presented at the top of 
         each subplot.\label{fig:stacks}}
\end{figure*}

Even though the DM of PSR J1745$-$1900 is as large as
$1778\ \mathrm{cm^{-3}\ pc}$, the integrated pulse profile of PSR
J1745$-$1900 above 2 GHz is dominated by pulse jitter rather than
scattering \citep{sle+14}.  Based on eight months GBT observations,
\citet{laks15} identified two periods of radio radiative activity for
PSR J1745$-$1900 : a stable state (covering MJDs 56515 - 56682) and an
erratic state (covering MJDs 56726 - 56845). In the stable state, the
evolution of the spin, radio flux density and profile shape remained
relatively stable, while in the erratic state, these properties varied
dramatically.

The mean pulse profiles and single-pulse stacks of PSR
J1745$-$1900 derived from the five Parkes observations are given in
Figure~\ref{fig:stacks}.  Apart from small variations discussed
below, the pulse profile remained steady
over the span of the observations, suggesting that the pulsar was in
the stable phase.  The observations presented here all occurred
prior to the stable state defined by \citet{laks15} (see
Table~\ref{tab:obs}).  This means that our results extend the
beginning of the stable state from MJD 56515 to MJD 56475, so the
stable state lasted for at least 6.8 months instead of the 5.5 months
reported by \citet{laks15}. 

The mean pulse profiles presented in Figure~\ref{fig:stacks} are
single peaked, consistent with early radio observations.  Pulse
profiles of PSR J1745$-$1900 obtained by early radio observations had
a single peak over a wide range of frequency \citep{sj13,sle+14}.
Later pulse profiles observed at the GBT at 8.7 GHz showed two peaks in
the stable state \citep{laks15}.  This led \citet{laks15} to suggest
that the pulse profile may have evolved from single to double peaked.

\begin{figure*}
\centering
\includegraphics[angle=0,width=0.8\textwidth]{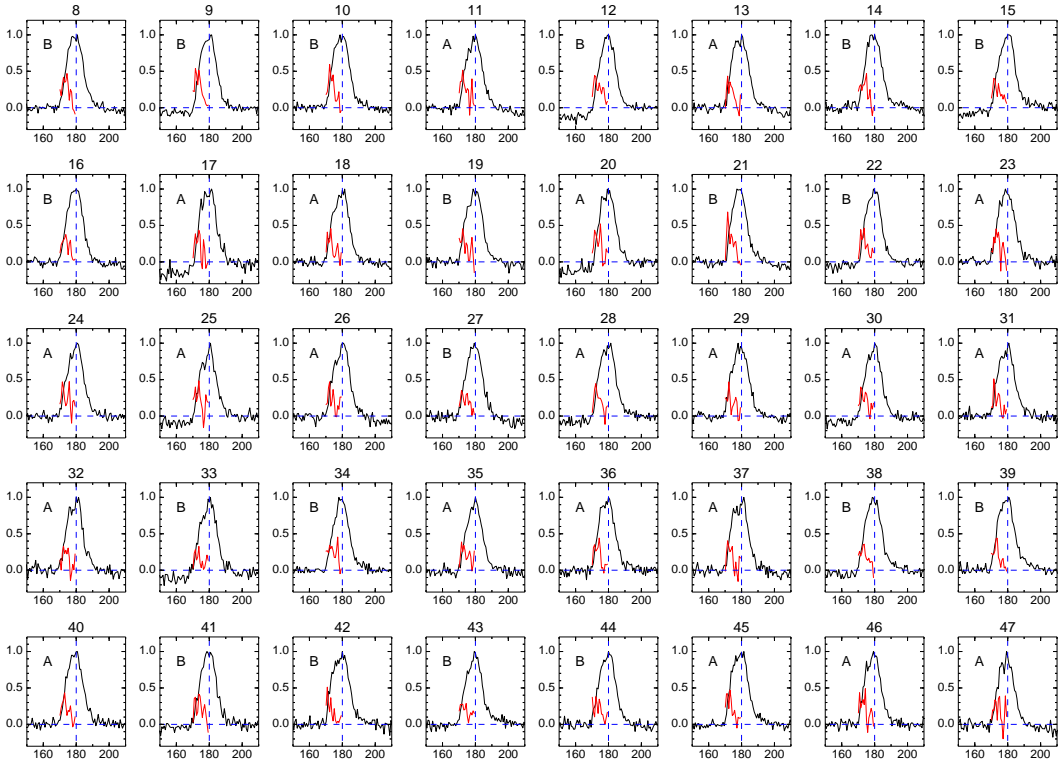}
\caption{A sequence of 40 sub-integration profiles of PSR J1745$-$2900
  with the sub-integration number on top, each averaged over 199.5
  seconds. The red line represents the slope of the leading
    edge for each sub-integration profile. The profile and slope zero
    are shown by the horizontal dashed line and the vertical dashed
    line represents the longitude of the peak of the integrated
    profile.
         \label{fig:mode_subint}}
\end{figure*}

\begin{figure}
\centering
\includegraphics[angle=0,width=0.4\textwidth]{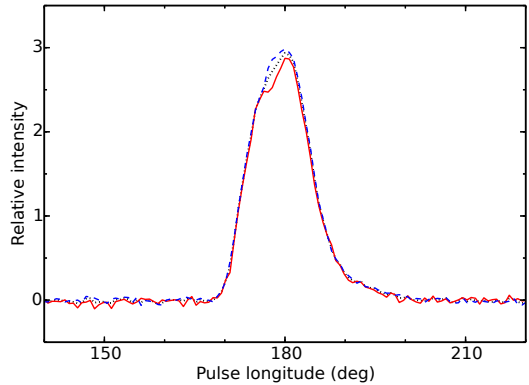}
\caption{Mean pulse profiles of J1745$-$2900 for mode A (solid), mode B 
        (dashed) and the two modes superimposed (dotted).
         \label{fig:mode_profile}}
\end{figure}

Although the mean pulse profile remains stable on a long time-scale
(on the order of several months) during the stable state, short
time-scale non-random pulse shape variations are visible in the single
pulse plots of Figure~\ref{fig:stacks}. To investigate the short term
pulse shape variations, high S/N sub-integration profiles are
needed. When the sub-integration time is less than 130 s, most
sub-integration profiles are dominated by pulse jitter and therefore
no systematic pulse shape variations can be seen.  By visual
inspection, we find that sub-integration profiles with sub-integration
times between 150~s to 250~s show similar systematic pulse shape
variations and so we chose a sub-integration time of 200~s (53
individual pulses) to study the short time-scale changes in pulse
shape. A selection of these sub-integration profiles is shown in
Figure~\ref{fig:mode_subint}. Some profiles (e.g. Nos. 24, 25 and 26)
show a strong main peak and a relatively weak but significant
secondary peak on the leading edge of the pulse profile, and we
classified these sub-integrations as mode A. Some sub-integration
profiles (e.g. Nos. 14, 15 and 16) show a single peak and we
classified these sub-integrations as mode B. To distinguish mode A and
mode B more rigorously, we used a method based on the point-to-point
slope of the leading edge to determine whether a profile has a
significant secondary peak on the leading edge or not. The
  slope at a given point is calculated by taking the average of the
  slopes between that point and its two closest neighbors.  If there
  is a significant secondary peak on the leading edge, indicating mode
  A, the slope will be negative at at least one point around the
  trailing edge of the secondary peak. If the slope does not go
  negative, the profile is designated as mode B. Using this method, we
  divided sub-integration profiles shown in
  Figure~\ref{fig:mode_subint} into the two modes.

  The duration of each mode is typically about ten minutes.
  However, we cannot obtain the exact duration of each mode because of
  S/N limitations.  Emission modes shorter than 200~s may exist and
  may be mixed with the other mode within a 200~s sub-integration.
  Sub-integration profiles Nos. 19, 33, 43 and 44 in
  Figure~\ref{fig:mode_subint} may have mixed modes.  There are hints
  of a secondary peak on the leading edge of these four profiles,
  but their secondary peaks are too weak to be reflected by the slope
  curve.

Mean pulse profiles for the two modes are given in
Figure~\ref{fig:mode_profile}. The mean pulse profile of mode A is
double peaked with a relatively weak leading peak while the mean pulse
profile of mode B is single peaked.

\subsection{Single pulses} \label{subsec:single}
Single-pulse properties of PSR J1745$-$1900 are presented in this subsection. 

\begin{figure*}
\centering
\includegraphics[angle=0,scale=1.0]{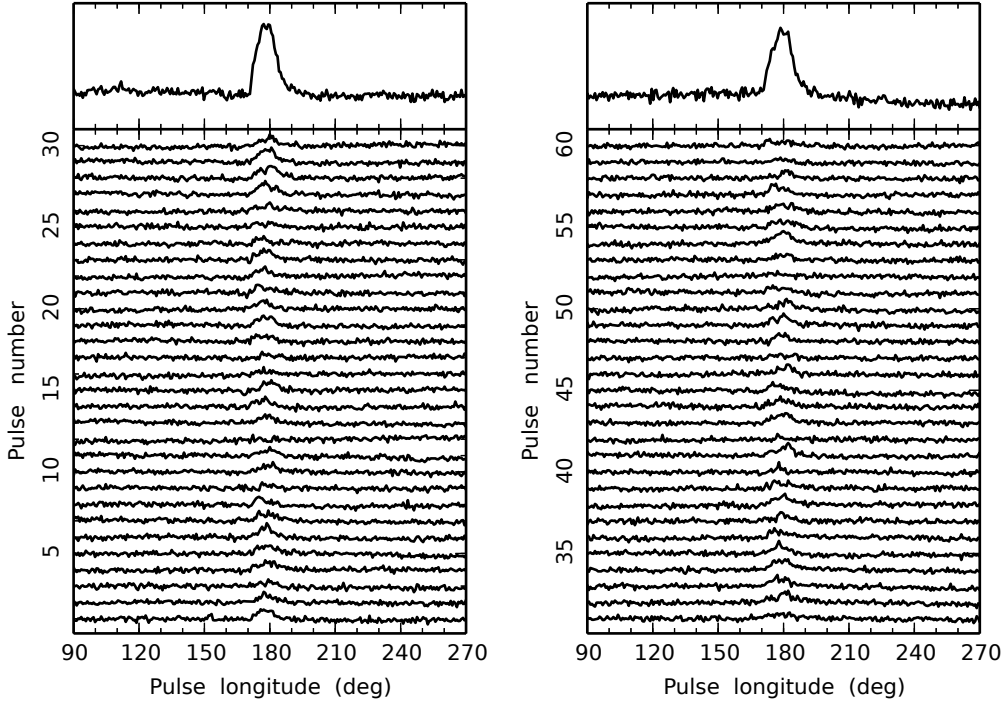}
\caption{Two contiguous sequences of successive individual pulses of 
         PSR J1745$-$2900 with their corresponding integrated 
         profiles at the top.\label{fig:seq}}
\end{figure*}

\begin{figure*}
\centering
\includegraphics[angle=0,scale=0.7]{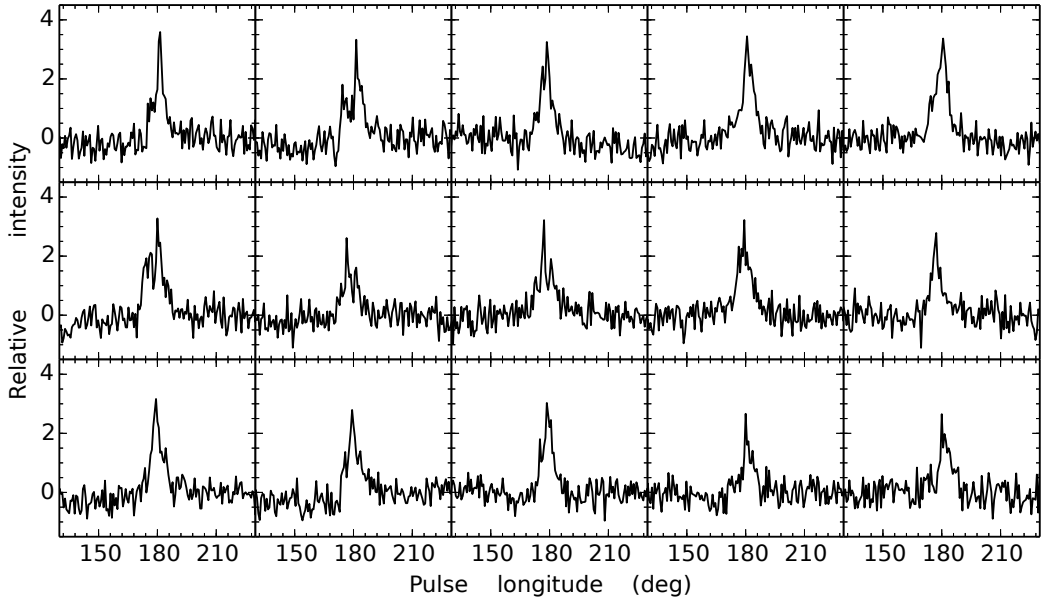}
\caption{Total intensity profiles for the fifteen highest peak S/N 
         individual pulses of PSR J1745$-$2900 observed on 
         MJD 56500.\label{fig:sample}}
\end{figure*}

Figure~\ref{fig:seq} shows two contiguous sequences of successive
individual pulses of PSR J1745$-$2900 and their corresponding
integrated profiles from the observation of MJD 56500. Each sequence
contains 30 consecutive single pulses. Single-pulse profiles of the
fifteen highest peak S/N pulses of PSR J1745$-$2900 observed on MJD
56500 are presented in Figure~\ref{fig:sample}. Similarly to
\citet{eam+12}, the peak S/N is calculated as the ratio between the
pulse peak amplitude in a given rotation period and the standard
deviation of the baseline points in the same period.

\subsubsection{Single-pulse energy distribution} \label{subsubsec:dist}

\begin{figure}
\centering
\includegraphics[angle=0,width=0.4\textwidth]{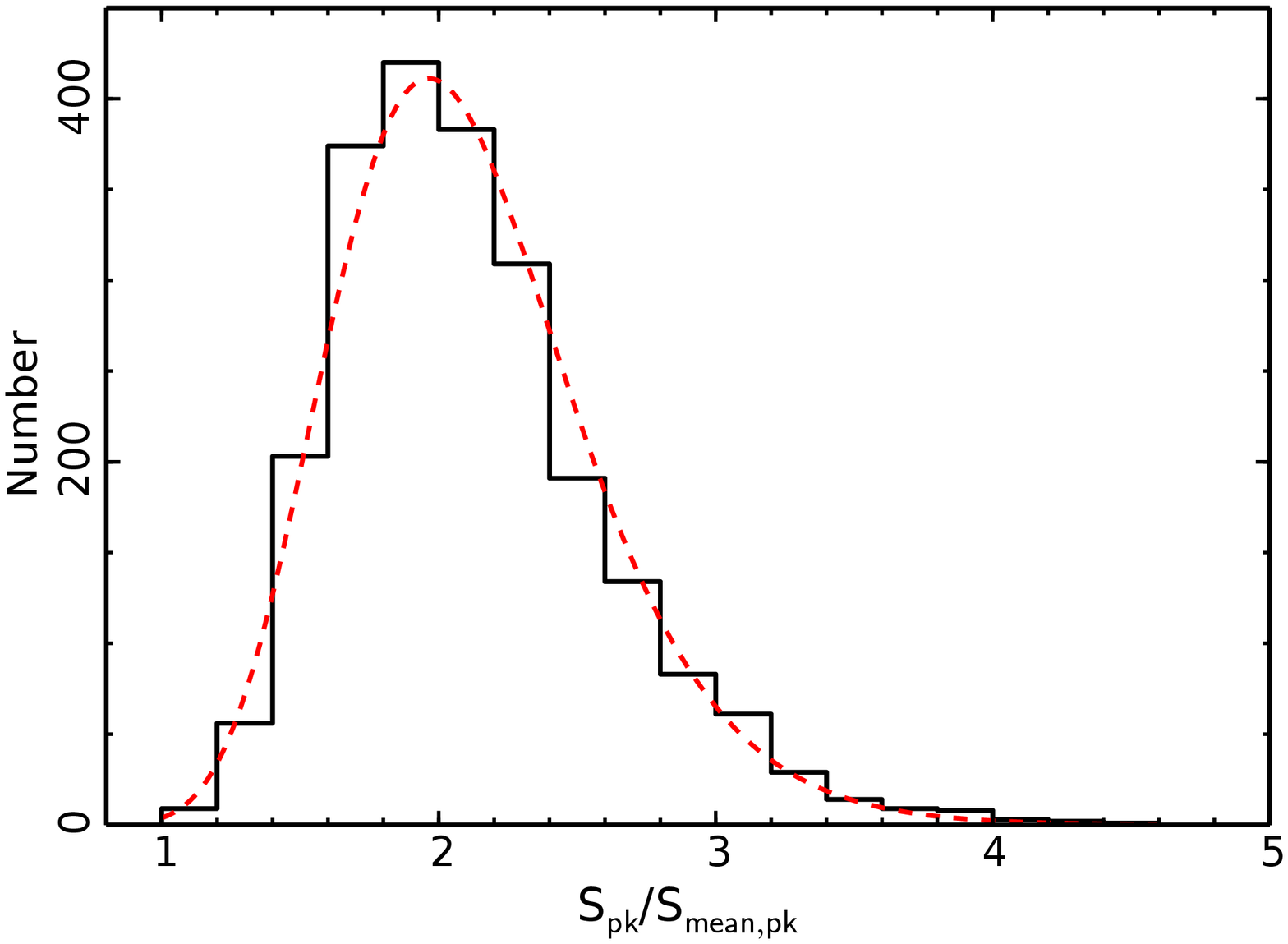}
\caption{Distribution of the peak flux density of pulses that have peak 
        S/N $\geq$ 4 for the five observations, normalized by 
        the peak flux density of the mean 
        pulse profile at that epoch. The dashed line represents 
         the best-fitting log-normal curve.\label{fig:spk}}
\end{figure}

\begin{figure}
\centering
\includegraphics[angle=0,width=0.4\textwidth]{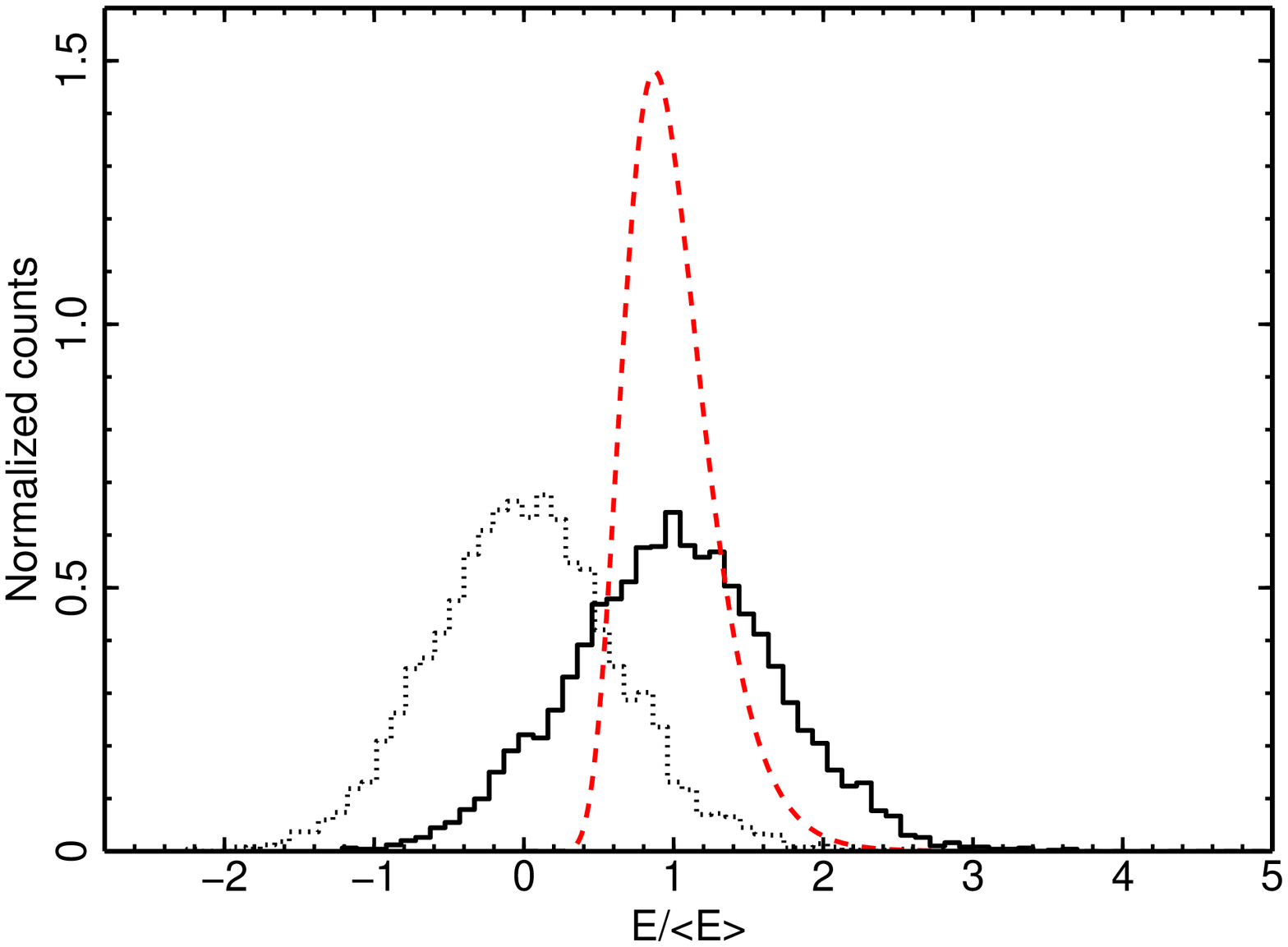}
\caption{Normalized pulse energy distributions for the five
  observations.  The solid, dotted and dashed lines represent the
  observed on-pulse energy distribution, the noise energy
  distribution, and the modelled intrinsic pulse energy distribution
  with a log-normal function, respectively.\label{fig:dist}}
\end{figure}

We present a statistical analysis of the energy distribution of single
pulses of the five observations.  The peak flux density distribution
of pulses with peak S/N $\geq$ 4 is presented in Figure~\ref{fig:spk},
and the normalized pulse energy distribution of all pulses is
presented in Figure~\ref{fig:dist}.

Following \citet{laks15} and \citet{ysw+15}, we normalized the peak flux 
density of each individual pulse, $\mathrm{S_{pk}}$, with the peak flux density 
of the mean pulse profile at corresponding observing epoch, $\mathrm{S_{mean,pk}}$. 
As we can see from Figure~\ref{fig:spk}, none of the single pulses 
has a peak flux density that is above 4.6 times the average and
there is therefore no evidence for giant pulses. Consistent with \citet{laks15} and 
\citet{ysw+15}, the peak flux density distribution 
shown in Figure~\ref{fig:spk} can be fitted well by a log-normal distribution. 
The log-normal probability density function is 
defined to be
\begin{equation}
P(\mathrm{E}/\left<\mathrm{E}\right>) = \frac{\left<\mathrm{E}\right>}{\sqrt{2\pi}\sigma \mathrm{E}}
\exp\left[-\left(\ln \frac{\mathrm{E}}{\left<\mathrm{E}\right>} -\mu\right)^2\middle/\left(2\sigma^2\right)\right],
\end{equation}
where $\mathrm{E}$ and $\left<\mathrm{E}\right>$ are the pulse energy of a
single pulse and the integrated pulse profile, respectively, and $\mu$
and $\sigma$ are the logarithmic mean and the standard deviation of
the distribution. 
In Figure~\ref{fig:spk}, 
the dashed line shows the best-fitting log-normal distribution with 
$\mu\ =\ 0.72$ and $\sigma\ =\ 0.22$. A Kolmogorov-Smirnov (KS) test was then 
performed and the $p$-value is $\sim$0.19, indicating that the fitted log-normal 
distribution is a good description for the peak flux density distribution.

Then we analyzed the pulse energy distribution of single pulses.  We
followed the procedure presented by \citet{wws+06} and \citet{wel16}
to model the observed pulse energy distribution by convolving an
intrinsic log-normal distribution with the observed noise distribution
for PSR J1745$-$2900.  The observed and the modelled intrinsic pulse
energy distributions are given in Figure~\ref{fig:dist}.  The observed
pulse energy was calculated by summing the intensities of the pulse
phase bins within the on-pulse region of the integrated pulse profile
at corresponding observing epoch.  The on-pulse window was
  defined as the total longitude range over which the pulse intensity
  significantly exceeds the baseline noise, that is, more than three
  times the baseline rms noise in several adjacent bins. The
observed noise energy was calculated in the same way using an equal
number of off-pulse bins.  Since the five observations were not flux
calibrated, we normalized the observed pulse energies and noise
energies with the pulse energy of the integrated pulse profile at
corresponding epoch. In Figure~\ref{fig:dist}, the dashed line shows
the modelled intrinsic log-normal distribution with $\mu\ =\ -0.05$
and $\sigma\ =\ 0.30$. Again a KS test was performed and the resulted
$p$-value is $\sim$0.77, showing that the intrinsic pulse energy
distribution is well described by the modelled log-normal distribution.

\subsubsection{Single-pulse widths} \label{subsubsec:width}

\begin{figure}
\centering
\includegraphics[angle=0,width=0.4\textwidth]{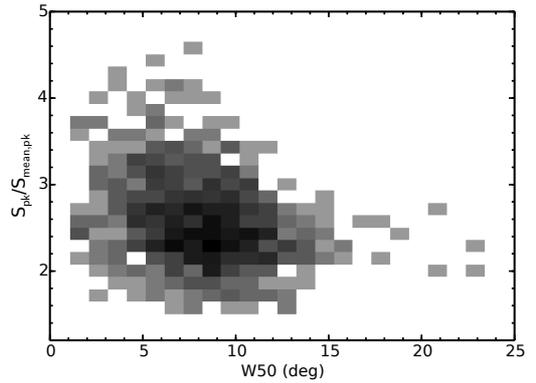}
\caption{Number density distribution of peak flux density of bright pulses relative
  to their half-power width W50 for the five
  observations.\label{fig:w50_spk}}
\end{figure}

\begin{figure}
\centering
\includegraphics[angle=0,width=0.4\textwidth]{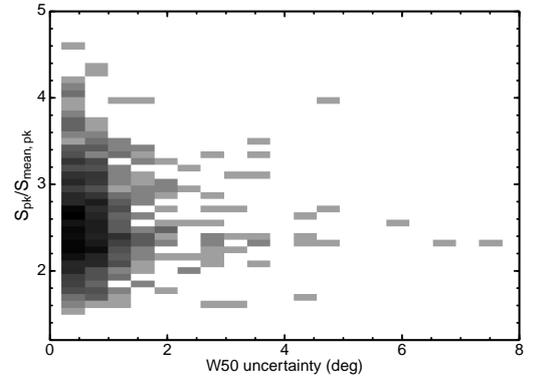}
\caption{Number density distribution of peak flux density of bright pulses relative
  to the uncertainty in their half-power width W50 for the five
  observations.\label{fig:w50_err}}
\end{figure}

Figure~\ref{fig:w50_spk} shows the two-dimensional distribution of
half-power pulse widths W50 versus peak flux density for PSR
J1745$-$2900 and Figure~\ref{fig:w50_err} shows the distribution of
uncertainties in the W50 measurements. Pulse widths were derived using
a linear interpolation between profile data points to define the pulse
phase at 50\% of the pulse peak. Uncertainties in the widths were
estimated by dividing the baseline rms noise level by the gradient of
the profile at each side and adding the width uncertainties for each
side in quadrature.

\begin{figure}
\centering
\includegraphics[angle=0,width=0.4\textwidth]{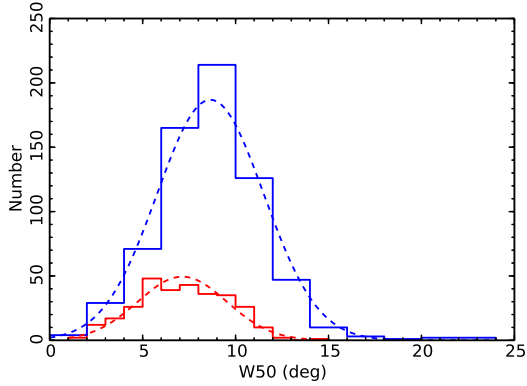}
\caption{Distributions of the pulse width W50 of weak pulses
    (S$\mathsf{_{pk}}$/S$\mathsf{_{mean,pk}}$ \textless 2.8, blue) and
    strong pulses (S$\mathsf{_{pk}}$/S$\mathsf{_{mean,pk}}$ $\ge$ 2.8,
    red). The dashed lines are the best-fitting Gaussian curves to
    each sample.
         \label{fig:w50_twobands}}
\end{figure}

  As has been observed in some other pulsars, e.g., PSR
  J0437$-$4715 \citep{jak+98} and the Crab pulsar PSR B0531+21
  \citep{mnl+11}, Figure~\ref{fig:w50_spk} shows an anticorrelation
  between peak flux density and pulse width, W50, for single
  pulses. Figure~\ref{fig:w50_err} shows that, apart from a few
  outliers, the width uncertainty for most single pulses is a degree
  or less, with no strong dependence on peak intensity. Consequently,
  the observed larger width of weaker pulses is not due to larger
  uncertainties. This conclusion is reinforced by
  Figure~\ref{fig:w50_twobands} which shows the distribution of widths
  averaged over two bands of peak flux density, respectively, pulses
  weaker than S$\mathsf{_{pk}}$/S$\mathsf{_{mean,pk}}$ $=$ 2.8,
  and greater than this value. The two distributions are well fitted
  by Gaussian curves and are signficantly different with mean widths
  of $8\fdg6$ for weaker pulses and $7\fdg1$ for stronger
  pulses. A two-sample KS test shows that the two Gaussian
  distributions are significantly different at the 95\% confidence
  level.

\begin{figure}
\centering
\includegraphics[angle=0,width=0.4\textwidth]{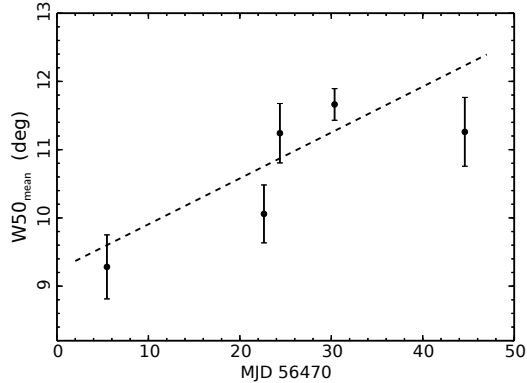}
\caption{Full width at half maximum of mean pulse profiles observed 
         at each epoch 
         as a function of time (points with error bars) 
         and the best linear fitting (dashed line) for 
         PSR J1745$-$2900.\label{fig:w50_time}}
\end{figure}

The time variations of the 50\% pulse width of mean pulse profiles
$\mathrm{W50_{mean}}$ are presented in Figure~\ref{fig:w50_time}.
Here, $\mathrm{W50_{mean}}$ is an average of half-power widths of each
subintegration of a given observation.  The plotted width uncertainties are
simply the standard deviation of the individual values used
to determine the average.  The results of \citet{laks15} showed that
there was an apparent increase in pulse profile width between MJDs
56544 and 56594 with a fitted rate of change of
$0\fdg08\ \pm\ 0\fdg04\ \mathrm{day^{-1}}$.  Our results shown in
Figure~\ref{fig:w50_time} are 
consistent with the results of \citet{laks15}, showing an obvious increase in
W50 between MJDs 56475 and 56514. The best-fit line gives a
slope of $0\fdg07\ \pm\ 0\fdg03\ \mathrm{day^{-1}}$.

\citet{ysw+15} reported that narrow spikes with half-power widths in
the range of $0\fdg2-0\fdg9$ were detected from this pulsar at 8.6
GHz. Their peak flux densities were at least 10 times larger than the
peak flux density of the mean pulse profile. In our results, the
half-power widths of the narrowest single pulses are about
$1\fdg1$.  However, their peak flux densities are no more than four
times of the peak flux density of the mean pulse profile.

\subsubsection{Subpulse drifting} \label{subsubsec:drifting}

\begin{figure}
\centering
\includegraphics[angle=-90,width=0.4\textwidth]{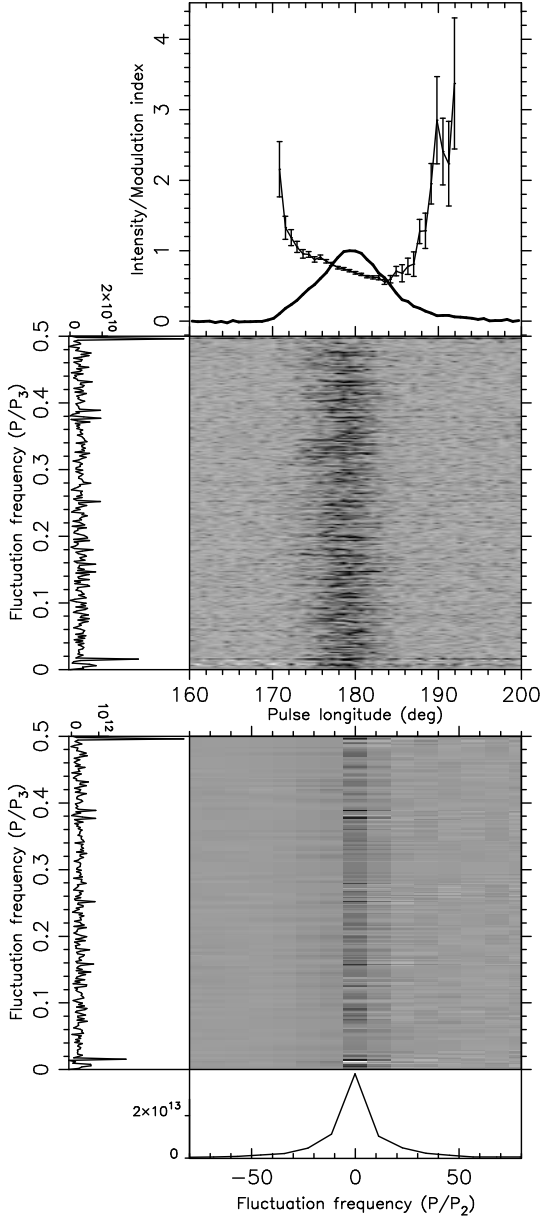}
\caption{Results of fluctuation analysis for the PSR J1745$-$2900 
         observation of MJD 56475. The top panel shows the mean pulse 
         profile (solid line) and longitude-resolved modulation index 
         (points with error bars). The LRFS and a side 
         panel showing the horizontally integrated power are given 
         below this panel. Below the LRFS, the 2DFS and  
         side panels showing horizontally (left) and vertically (bottom) 
         integrated power are plotted.\label{fig:lrfs}}
\end{figure}

Many pulsars show the phenomenon of subpulse drifting in which 
the subpulses drift in pulse phase or longitude across a sequence 
of single pulses (e.g. \citealt{wes06}). The subpulse drifting pattern 
is quasi-periodic with a characteristic spacing
of the subpulses in pulse longitude (P2) and pulse number (P3). 
In order to investigate the variability of subpulses, we carried out 
an analysis of fluctuation spectra by calculating the longitude-resolved 
modulation index, the longitude-resolved fluctuation spectrum (LRFS) 
\citep{bac70b} 
and the two-dimensional fluctuation spectrum (2DFS) \citep{es02} 
for the five observations. The longitude-resolved modulation index is a measure 
of the amount of intensity variability at a given pulse longitude, 
while the LRFS and the 2DFS are used to characterize P3 and P2 respectively.
For more details about the techniques of analysis, we refer to \citet{wes06}. 
Figure~\ref{fig:lrfs} gives an example of our results based on the observation 
of MJD 56475. The asymmetric distribution of modulation index presented in the 
top panel indicates that 
the intensity variation is different between the leading and trailing edge of 
the mean pulse profile. Previously, \citet{laks15} and \citet{ysw+15} 
searched for drifting subpulses in their observations but no evidence was 
found for the existence of drifting subpulses. In Figure~\ref{fig:lrfs}, the 
side panels of the LRFS and 2DFS show spectral features at $P/P_3\ \simeq\ 0.02$ 
cycles per period (cpp) 
and $P/P_3\ \simeq\ 0.5$ cpp. However, these spectral features are both produced by 
interference because they are visible at the whole range of pulse longitude. 
No subpulse modulation feature can be seen in either the LRFS or the 2DFS. 
This confirms the results presented by \citet{laks15} and \citet{ysw+15} 
that subpulse drifting is not detectable in PSR J1745$-$2900. 

\subsection{Pulse nulling} \label{subsec:nulling}

\begin{figure*}
\centering
\includegraphics[angle=0,width=0.7\textwidth]{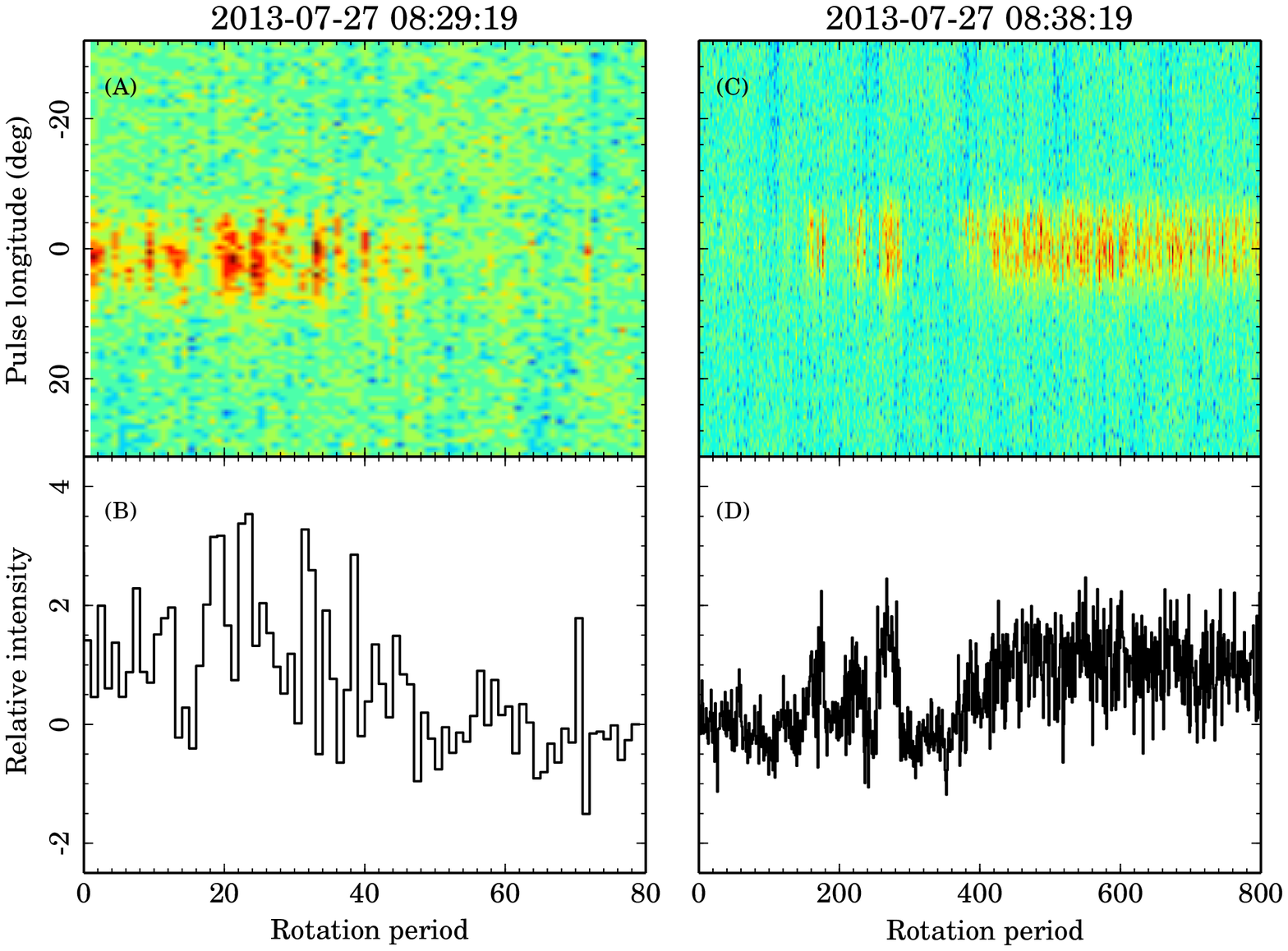}
\caption{Pulse energy variations with time for the observation of 
         MJD 56500 (panel (D)) 
         and a 5-min prior observation (panel (B)) with corresponding single-pulse 
         stacks on top (panels (A) and (C)). The start UTC times of 
         the two observations are shown on the top of each column. 
         \label{fig:nulling}}
\end{figure*}

\begin{figure}
\centering
\includegraphics[angle=0,width=0.4\textwidth]{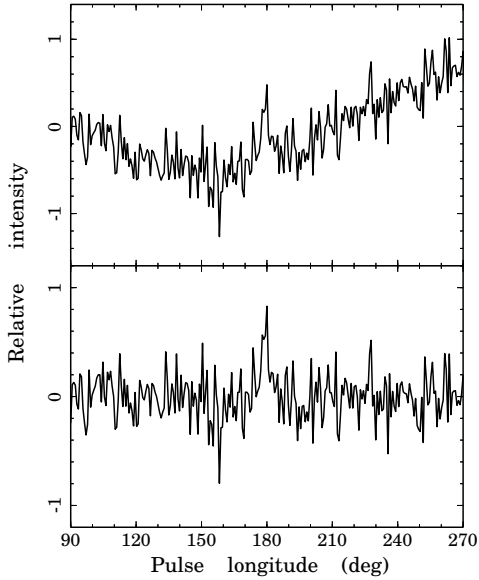}
\caption{Mean pulse profile of null pulses between rotation period 80 and 100 
         in the right column of Figure~\ref{fig:nulling} (upper) and the same profile 
         after baseline fluctuation mitigation (lower).}
         \label{fig:nulling_profile}
\end{figure}

\begin{figure}
\centering
\includegraphics[angle=0,width=0.4\textwidth]{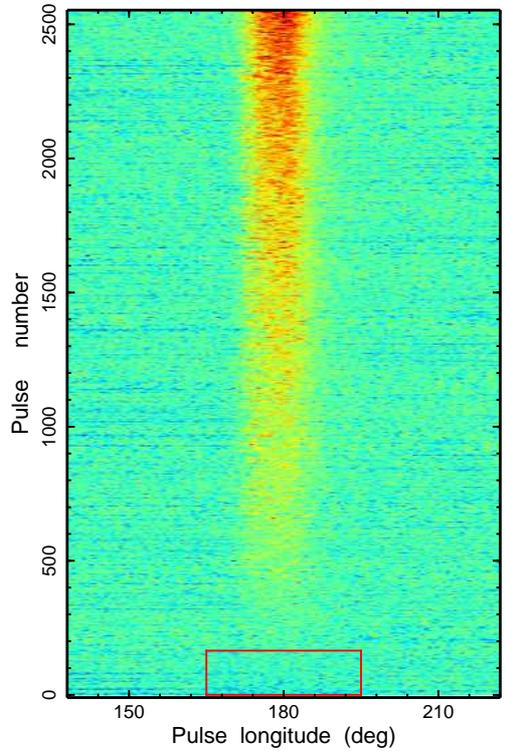}
\caption{Arranged individual pulses in the ascending order of their on-pulse 
         energy for the observation of MJD 56500. The null pulses can 
         be seen towards the low energy end (inside an indicative red box).
         \label{fig:nulling_rearr}}
\end{figure}

\begin{figure}
\centering
\includegraphics[angle=0,width=0.4\textwidth]{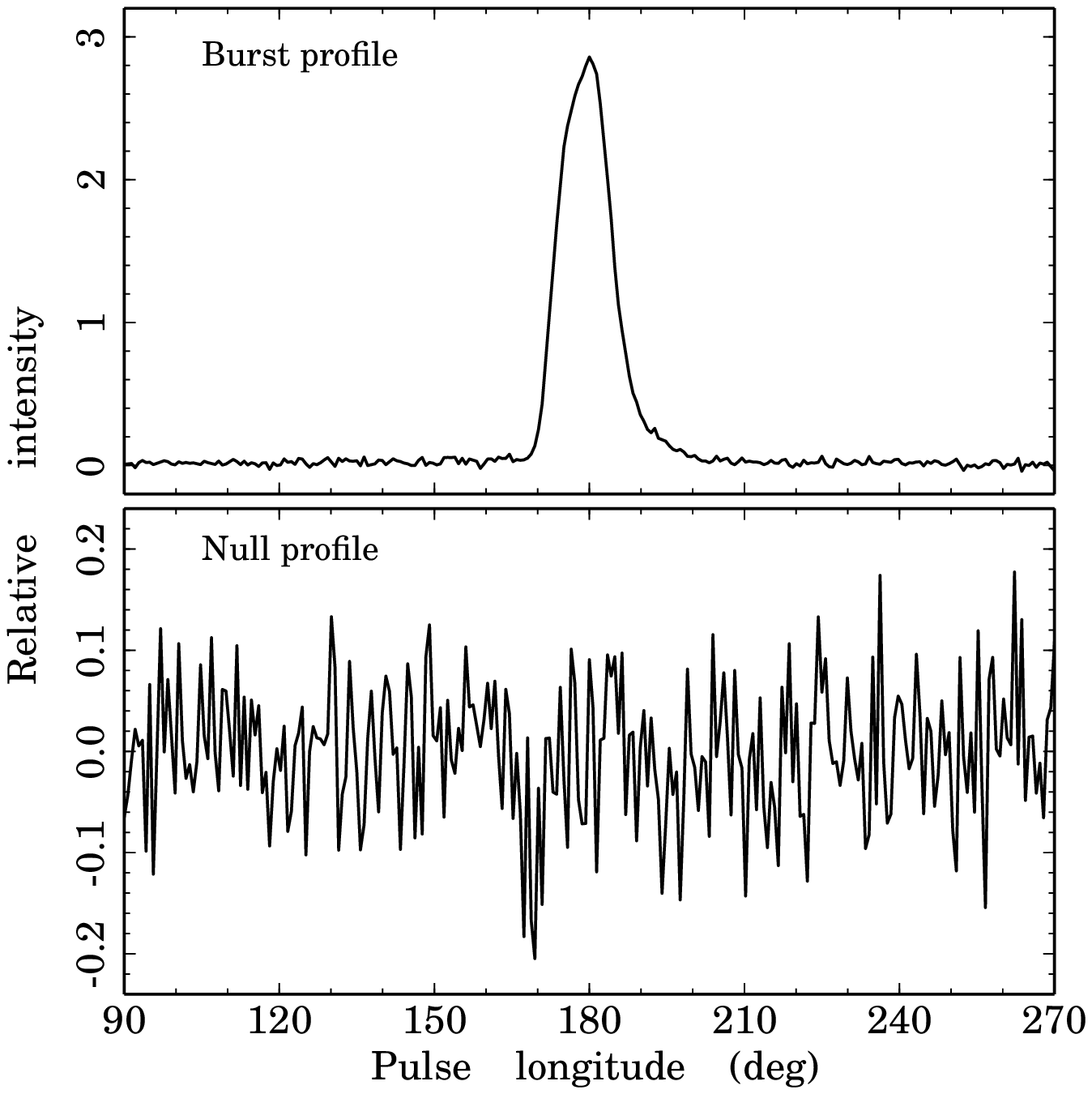}
\caption{Mean pulse profiles of null pulses (lower)
        and burst pulses (upper).
         \label{fig:weak}}
\end{figure}

Pulsar nulling is a phenomenon in which pulsed emission suddenly turns off 
for several pulse periods and then just as suddenly turns on 
(e.g. \citealt{wmj07}). Nulling is relatively common in pulsars and observed 
mostly in longer-period pulsars. Nulling has been shown to occur in more than 
100 pulsars to date \citep{rit76,ran86,big92a,viv95,wmj07,bjb+12,gjk12}. 
Although the radio emission properties of magnetars are similar to those of 
the normal pulsars in several respects \citep{ksj+07}, nulling has not been 
reported in magnetars before. Visual inspection of Figure~\ref{fig:stacks} 
shows that there are several apparent nulls in the observation of MJD 56500. 
The pulse energy distribution shown in Figure~\ref{fig:dist} has a
small secondary peak at $E/\langle E\rangle = 0$, also suggesting pulse nulling or 
weak modes in PSR J1745$-$2900. 

The pulse energy variations with time are presented in
Figure~\ref{fig:nulling} for the observation of MJD 56500 (right
column) and a short 5-min observation 9 minutes earlier (left column).
The upper panels show the color-scale plots of single-pulse
intensities for the two observations.  Corresponding pulse energy
variations are shown in the lower panels.  As we can see in panels (A)
and (B), the pulsar fades out after rotation period 44.  Panel (C)
shows that there are obvious pulse cessations in period ranges of
$\sim$1-150, 180-220, 230-250 and 300-370. These cessations look very
like pulsar nulling.  Panel (D) shows clearly that the pulse energy
drops to zero several times before the rotation period 370 and return
to normal after then. This is consistent with the nulling phenomenon
observed in normal pulsars. Since the gap between the
  UTC 08:29:19 observation and the UTC 08:38:19 observation is about four
  minutes, panels (B) and (D) suggest that the nulling event lasted
  about 30 minutes (from rotation period 44 in panel (B) to rotation
  period 370 in panel (D)). Of course it is also possible that the
  pulsed emission switched on again during the observation gap.

Weak emission was detected in PSR B0826$-$34 during an apparent null
phase \citep{elg+05}. We therefore investigated whether the apparent
nulls seen in PSR J1745$-$2900 are real nulls or weak modes by forming
null-phase pulse profile by averaging the pulses in the null
state. For simplicity, we chose pulses in the most apparent null
state, i.e., pulses between rotation period 80 and 100 in
Figure~\ref{fig:nulling}, to form the mean pulse profile. The
  profile obtained from these null pulses is shown in the upper panel
  of Figure~\ref{fig:nulling_profile}.  Baseline fluctuations are
  present in our data. These can arise from receiver fluctuations,
  atmospheric fluctuations or they may be intrinsic to the
  pulsar. Similar baseline fluctuations were seen by \citet{laks15}
  and attributed by them to changes in atmospheric opacity.
  Simultaneous observations at multiple telescopes and frequencies
  would help to establish the origin of these fluctuations. We tried
  to mitigate their effect in our data by fitting a sine function to
  the baseline and subtracting it from the data. The resulted profile
  is given in the lower panel. Surprisingly, the average profile of
the null pulses shows a clear detection of the pulse profile.  It is
therefore important to know if the weak emission profile arises from
bright but rare single pulses, or if it is a steady weak emission.

Following \citet{gyy+17}, to investigate this we arranged all
  pulses from the MJD 56500 observation in ascending order of
  their on-pulse energy as shown in Figure~\ref{fig:nulling_rearr}.
If the weak emission profile of null pulses originates from a few
single pulses, there should be no evidence of any emission when we
just average several null pulses with lowest on-pulse energy. On the
contrary, if the weak emission profile originates from intrinsic weak
emission, there should be evidence of emission even if we only average
a few weak null pulses. We formed mean pulse profiles by
  averaging different numbers of the weakest null pulses as indicated
  by the box at the bottom of Figure~\ref{fig:nulling_rearr}. The
  upper edge of the box was moved toward the high-energy end until the
  mean profile for pulses inside the box show a significant emission
  profile (peak S/N $> 3\sigma$).  This level was used to divide all
  pulses into two groups with pulses outside this box marked as burst
  pulses and the pulses inside the box marked as null pulses.

After this separation, the average profiles obtained from the null pulses 
and the burst pulses are presented in Figure~\ref{fig:weak}. The mean pulse profile 
of null pulses in the lower panel does not show any significant emission component.
This suggests that the weak emission profile shown in Figure~\ref{fig:nulling_profile} 
arises from bright but rare single pulses instead of a steady weak emission.
Hence we can conclude that the apparent null states observed on MJD 56500 
are real nulls rather than being a weak-emission mode.

\section{Discussion and conclusions} \label{sec:discn}

We have presented the mean pulse profile and single-pulse properties
of PSR J1745$-$2900 at 3.1 GHz by analyzing five observations with
high S/N made at epochs between 2013 July and August.  The data were
downloaded from the Parkes Pulsar Data Archive.

A stable radio state is not often seen in radio-emitting
magnetars. During the stable state, the magnetar acts more like a
normal pulsar, in that both the radio pulse profile shape and flux
density are stable.  The similarity of the mean pulse profile shape of
the five observations indicates that the pulsar was in a stable state
between MJDs 56475 and 56514.  As the observations analyzed here
occurred before the GBT observations reported by \citet{laks15}, the
observations extend the stable state of this pulsar from 5.5 months to
6.8 months.

In spite of the large DM, interstellar scattering is not a
  dominating effect for the mean pulse profile of PSR J1745$-$2900 at
  3.1 GHz with the expected scattering timescale being about 18~ms or
  $1\fdg7$ of pulse phase \citep{sle+14}.  We detected a linear
  variation with a slope of $0\fdg07\ \pm\ 0\fdg03\ \mathrm{day^{-1}}$
  in the pulse width of mean pulse profiles between MJDs 56544 and
  56594, which confirms and extends the pulse width variations
  presented by \citet{laks15}.

\citet{laks15} noted that the pulse profile prior to their GBT
  observations was single peaked while the pulse profile in the stable
  state of their observations was double peaked. Similar to the pulse
  width evolution, they found that the component separation increased
  with time during their observations, with a value of $5\fdg5$ ($\sim$57 ms) at
  the time of their first observation on MJD 56515. The last Parkes
  observation was just one day earlier at MJD 56514 and, for all of
  our observations, the pulse profile was single-peaked
  (Figure~\ref{fig:stacks}). Since the \citet{sle+14} 3.2~GHz
  observations were almost coincident with those at Parkes, it is
  unlikely that the single-peaked Parkes profiles result from scatter
  broadening. In view of the near coincidence in time of the last
  Parkes observation and the first GBT observation, it seems more
  probable that at this time there was an evolution in frequency of the
  shape of the mean pulse profile, with the components becoming better
  defined at higher frequencies.

Single-pulse observations of some pulsars, for example, PSR B0656+14
\citep{wws+06}, PSR B1839$-$04 \citep{wel16} and PSR J1713+0747
\citep{lbj+16}, showed that the observed on-pulse energy distribution
can be modelled by convolving an intrinsic distribution with the
observed noise distribution. With the same analysis, we found that the
intrinsic pulse energy distribution of PSR J1745$-$2900 at 3.1 GHz is
well described by a log-normal distribution. We found an
  anticorrelation between peak flux density of single pulses and their
  50\% width similar to those observed in the millisecond pulsar PSR
  J0437$-$4715 \citep{jak+98} and the Crab pulsar PSR B0531+21
  \citep{mnl+11}. Stronger pulses have a mean width of $7\fdg1$ and
  for weaker pulses the mean width is $8\fdg6$. We showed that the
  pulse width distributions for strong and weak single pulses are
  significantly different at the 95\% confidence level.  Consistent
with the results presented by \citet{laks15} and \citet{ysw+15}, giant
pulses and subpulse drifting were not detected in the five
observations for J1745$-$2900 at 3.1 GHz.

Besides the long time-scale evolution of the mean pulse profile, PSR
J1745$-$2900 also shows short time-scale pulse shape variations. By
forming short sub-integrations, we found evidence for mode changing on
timescales of several minutes. One mode has two clear overlapping
components whereas the other only shows a broad single component. 

We detected pulsar nulling in PSR J1745$-$2900.
In the observation of MJD 56500, the pulse energy drops to
zero several times then return to the normal level. We could find no
evidence for instrumental problems that could cause apparent nulls and
neither diffractive nor refractive scintillation can account for the
variations. The observing band covers many diffractive bands and the
timescale for refractive scintillation is of the order years
\citep{bdd+15,ppe+15}. Summing of data within null regions reveals a
weak pulse, but we believe that this results from an occasional strong
pulse rather than indicating that the null is a weak emission mode. 

In some respects, the radio emission of magnetars is different from
that of normal radio pulsars.  For magnetars, the radio emission is
transient, the radio flux density and the pulse profile are highly
variable, and the radio spectrum is relatively flat.  The radio
emission of XTE J1810$-$197 (PSR J1809$-$1943) is always extremely
variable, its radio flux density and pulse shape showed dramatic
changes on time-scales ranging from minutes to weeks
\citep{crj+07,ksj+07,ljk+08,crh+16}. The flux density of PSR
J1622$-$4950 varies up to a factor of $\sim$10 within a few days and
the observed pulse shapes of PSR J1622$-$4950 changes significantly
from day to day \citep{lbb+10,lbb+12}. PSR J1550$-$5418 also showed
variations in pulse profile shape and flux density, with the flux
density varying by factors up to $\sim$50$\%$ on timescales of a few
days \citep{crj+08}.  These variations observed in radio-emitting
magnetars are intrinsic to the pulsar and may be related to changes in
magnetospheric plasma densities and/or currents. Based on the
differences between magnetar radio emission and normal pulsar radio
emission, it has been proposed that the radio emission of
magnetars is powered by magnetic field decay instead of by rotation
\citep{tyl13}. However, in other respects the radio emission
  from the two classes of pulsar is similar. For example, the pulse
  polarization properties of magnetars are similar to those of other
  pulsars \citep[e.g.,][]{crj+07,efk+13}. The presence of mode changing and
  pulsar nulling in the GC magnetar PSR J1745$-$2900 gives further
  support to the idea that the radio emission mechanism is bascially
  the same in magnetars and normal pulsars.

\section*{Acknowledgements}

This work is supported by National Basic Research Program
of China (973 Program 2015CB857100), 
West Light Foundation of Chinese Academy of Sciences 
(No. XBBS201422), National Natural Science Foundation of China (Nos. U1631106, 
U1731238), 
the Strategic Priority Research Program of Chinese Academy of Sciences 
(No. XDB23010200) and the National Key Research and Development Program of China 
(No. 2016YFA0400800). ZGW acknowledges support from West light Foundation of CAS 
(2016-QNXZ-B-24). 
We thank an anonymous referee for helpful comments that
improved the manuscript. 
We thank K. J. Lee, V. Gajjar, R. Yuen and J. M. Yao for 
valuable discussions. We also thank P. Weltevrede for suggestions on 
the usage of the {\tt\string PSRSALSA} package. 
The Parkes radio telescope is part of the 
Australia Telescope, which is funded by the Commonwealth of Australia for 
operation as a National Facility managed by the Commonwealth Scientific 
and Industrial Research Organisation.

\bibliographystyle{mnras}


\begin{thebibliography}{}
\makeatletter
\relax
\def\mn@urlcharsother{\let\do\@makeother \do\$\do\&\do\#\do\^\do\_\do\%\do\~}
\def\mn@doi{\begingroup\mn@urlcharsother \@ifnextchar [ {\mn@doi@}
  {\mn@doi@[]}}
\def\mn@doi@[#1]#2{\def\@tempa{#1}\ifx\@tempa\@empty \href
  {http://dx.doi.org/#2} {doi:#2}\else \href {http://dx.doi.org/#2} {#1}\fi
  \endgroup}
\def\mn@eprint#1#2{\mn@eprint@#1:#2::\@nil}
\def\mn@eprint@arXiv#1{\href {http://arxiv.org/abs/#1} {{\tt arXiv:#1}}}
\def\mn@eprint@dblp#1{\href {http://dblp.uni-trier.de/rec/bibtex/#1.xml}
  {dblp:#1}}
\def\mn@eprint@#1:#2:#3:#4\@nil{\def\@tempa {#1}\def\@tempb {#2}\def\@tempc
  {#3}\ifx \@tempc \@empty \let \@tempc \@tempb \let \@tempb \@tempa \fi \ifx
  \@tempb \@empty \def\@tempb {arXiv}\fi \@ifundefined
  {mn@eprint@\@tempb}{\@tempb:\@tempc}{\expandafter \expandafter \csname
  mn@eprint@\@tempb\endcsname \expandafter{\@tempc}}}

\bibitem[\protect\citeauthoryear{Backer}{Backer}{1970}]{bac70b}
Backer D.~C.,  1970, Nature, 227, 692

\bibitem[\protect\citeauthoryear{Biggs}{Biggs}{1992}]{big92a}
Biggs J.~D.,  1992, ApJ, 394, 574

\bibitem[\protect\citeauthoryear{{Bower} et~al.,}{{Bower}
  et~al.}{2014}]{bdd+14}
{Bower} G.~C.,  et~al., 2014, \mn@doi [ApJ] {10.1088/2041-8205/780/1/L2}, \href
  {http://adsabs.harvard.edu/abs/2014ApJ...780L...2B} {780, L2}

\bibitem[\protect\citeauthoryear{{Bower} et~al.,}{{Bower}
  et~al.}{2015}]{bdd+15}
{Bower} G.~C.,  et~al., 2015, \mn@doi [ApJ] {10.1088/0004-637X/798/2/120},
  \href {http://adsabs.harvard.edu/abs/2015ApJ...798..120B} {798, 120}

\bibitem[\protect\citeauthoryear{{Burke-Spolaor} et~al.,}{{Burke-Spolaor}
  et~al.}{2012}]{bjb+12}
{Burke-Spolaor} S.,  et~al., 2012, \mn@doi [MNRAS]
  {10.1111/j.1365-2966.2012.20998.x}, \href
  {http://adsabs.harvard.edu/abs/2012MNRAS.423.1351B} {423, 1351}

\bibitem[\protect\citeauthoryear{{Camilo}, {Ransom}, {Halpern}, {Reynolds},
  {Helfand}, {Zimmerman}  \& {Sarkissian}}{{Camilo} et~al.}{2006}]{crh+06}
{Camilo} F.,  {Ransom} S.~M.,  {Halpern} J.~P.,  {Reynolds} J.,  {Helfand}
  D.~J.,  {Zimmerman} N.,   {Sarkissian} J.,  2006, \mn@doi [Nature]
  {10.1038/nature04986}, 442, 892

\bibitem[\protect\citeauthoryear{{Camilo}, {Reynolds}, {Johnston}, {Halpern},
  {Ransom}  \& {van Straten}}{{Camilo} et~al.}{2007a}]{crj+07}
{Camilo} F.,  {Reynolds} J.,  {Johnston} S.,  {Halpern} J.~P.,  {Ransom} S.~M.,
    {van Straten} W.,  2007a, \mn@doi [ApJ] {10.1086/516630}, 659, L37

\bibitem[\protect\citeauthoryear{{Camilo}, {Ransom}, {Halpern}  \&
  {Reynolds}}{{Camilo} et~al.}{2007b}]{crhr07}
{Camilo} F.,  {Ransom} S.~M.,  {Halpern} J.~P.,   {Reynolds} J.,  2007b,
  \mn@doi [ApJ] {10.1086/521826}, 666, L93

\bibitem[\protect\citeauthoryear{{Camilo}, {Reynolds}, {Johnston}, {Halpern}
  \& {Ransom}}{{Camilo} et~al.}{2008}]{crj+08}
{Camilo} F.,  {Reynolds} J.,  {Johnston} S.,  {Halpern} J.~P.,   {Ransom}
  S.~M.,  2008, \mn@doi [ApJ] {10.1086/587054}, \href
  {http://adsabs.harvard.edu/abs/2008ApJ...679..681C} {679, 681}

\bibitem[\protect\citeauthoryear{{Camilo} et~al.,}{{Camilo}
  et~al.}{2016}]{crh+16}
{Camilo} F.,  et~al., 2016, \mn@doi [ApJ] {10.3847/0004-637X/820/2/110}, \href
  {http://adsabs.harvard.edu/abs/2016ApJ...820..110C} {820, 110}

\bibitem[\protect\citeauthoryear{Duncan \& Thompson}{Duncan \&
  Thompson}{1992}]{dt92a}
Duncan R.~C.,  Thompson C.,  1992, ApJ, 392, L9

\bibitem[\protect\citeauthoryear{{Eatough} et~al.,}{{Eatough}
  et~al.}{2013}]{efk+13}
{Eatough} R.~P.,  et~al., 2013, \mn@doi [Nature] {10.1038/nature12499}, \href
  {http://adsabs.harvard.edu/abs/2013Natur.501..391E} {501, 391}

\bibitem[\protect\citeauthoryear{{Edwards} \& {Stappers}}{{Edwards} \&
  {Stappers}}{2002}]{es02}
{Edwards} R.~T.,  {Stappers} B.~W.,  2002, A\&A, 393, 733

\bibitem[\protect\citeauthoryear{{Esamdin}, {Lyne}, {Graham-Smith}, {Kramer},
  {Manchester}  \& {Wu}}{{Esamdin} et~al.}{2005}]{elg+05}
{Esamdin} A.,  {Lyne} A.~G.,  {Graham-Smith} F.,  {Kramer} M.,  {Manchester}
  R.~N.,   {Wu} X.,  2005, MNRAS, 356, 59

\bibitem[\protect\citeauthoryear{{Esamdin}, {Abdurixit}, {Manchester}  \&
  {Niu}}{{Esamdin} et~al.}{2012}]{eam+12}
{Esamdin} A.,  {Abdurixit} D.,  {Manchester} R.~N.,   {Niu} H.~B.,  2012,
  \mn@doi [ApJ] {10.1088/2041-8205/759/1/L3}, \href
  {http://ads.bao.ac.cn/abs/2012ApJ...759L...3E} {759, L3}

\bibitem[\protect\citeauthoryear{{Gajjar}, {Joshi}  \& {Kramer}}{{Gajjar}
  et~al.}{2012}]{gjk12}
{Gajjar} V.,  {Joshi} B.~C.,   {Kramer} M.,  2012, \mn@doi [MNRAS]
  {10.1111/j.1365-2966.2012.21296.x}, \href
  {http://adsabs.harvard.edu/abs/2012MNRAS.424.1197G} {424, 1197}

\bibitem[\protect\citeauthoryear{{Gajjar}, {Yuan}, {Yuen}, {Wen}, {Liu}  \&
  {Wang}}{{Gajjar} et~al.}{2017}]{gyy+17}
{Gajjar} V.,  {Yuan} J.~P.,  {Yuen} R.,  {Wen} Z.~G.,  {Liu} Z.~Y.,   {Wang}
  N.,  2017, \mn@doi [ApJ] {10.3847/1538-4357/aa96ac}, \href
  {http://adsabs.harvard.edu/abs/2017ApJ...850..173G} {850, 173}

\bibitem[\protect\citeauthoryear{{Hobbs} et~al.,}{{Hobbs}
  et~al.}{2011}]{hmm+11}
{Hobbs} G.,  et~al., 2011, \mn@doi [PASA] {10.1071/AS11016}, \href
  {http://adsabs.harvard.edu/abs/2011PASA...28..202H} {28, 202}

\bibitem[\protect\citeauthoryear{{Hotan}, {van Straten}  \&
  {Manchester}}{{Hotan} et~al.}{2004}]{hvm04}
{Hotan} A.~W.,  {van Straten} W.,   {Manchester} R.~N.,  2004, PASA, 21, 302

\bibitem[\protect\citeauthoryear{Jenet, Anderson, Kaspi, Prince  \&
  Unwin}{Jenet et~al.}{1998}]{jak+98}
Jenet F.,  Anderson S.,  Kaspi V.,  Prince T.,   Unwin S.,  1998, ApJ, 498, 365

\bibitem[\protect\citeauthoryear{{Kennea} et~al.,}{{Kennea}
  et~al.}{2013}]{kbk+13}
{Kennea} J.~A.,  et~al., 2013, \mn@doi [ApJ] {10.1088/2041-8205/770/2/L24},
  \href {http://adsabs.harvard.edu/abs/2013ApJ...770L..24K} {770, L24}

\bibitem[\protect\citeauthoryear{{Kramer}, {Stappers}, {Jessner}, {Lyne}  \&
  {Jordan}}{{Kramer} et~al.}{2007}]{ksj+07}
{Kramer} M.,  {Stappers} B.~W.,  {Jessner} A.,  {Lyne} A.~G.,   {Jordan} C.~A.,
   2007, MNRAS, 377, 107

\bibitem[\protect\citeauthoryear{{Krishnakumar}, {Mitra}, {Naidu}, {Joshi}  \&
  {Manoharan}}{{Krishnakumar} et~al.}{2015}]{kmn+15}
{Krishnakumar} M.~A.,  {Mitra} D.,  {Naidu} A.,  {Joshi} B.~C.,   {Manoharan}
  P.~K.,  2015, \mn@doi [ApJ] {10.1088/0004-637X/804/1/23}, \href
  {http://adsabs.harvard.edu/abs/2015ApJ...804...23K} {804, 23}

\bibitem[\protect\citeauthoryear{{Lazaridis}, {Jessner}, {Kramer}, {Stappers},
  {Lyne}, {Jordan}, {Serylak}  \& {Zensus}}{{Lazaridis} et~al.}{2008}]{ljk+08}
{Lazaridis} K.,  {Jessner} A.,  {Kramer} M.,  {Stappers} B.~W.,  {Lyne} A.~G.,
  {Jordan} C.~A.,  {Serylak} M.,   {Zensus} J.~A.,  2008, \mn@doi [MNRAS]
  {10.1111/j.1365-2966.2008.13794.x}, \href
  {http://adsabs.harvard.edu/abs/2008MNRAS.390..839L} {390, 839}

\bibitem[\protect\citeauthoryear{{Levin} et~al.,}{{Levin}
  et~al.}{2010}]{lbb+10}
{Levin} L.,  et~al., 2010, \mn@doi [ApJ] {10.1088/2041-8205/721/1/L33}, \href
  {http://adsabs.harvard.edu/abs/2010ApJ...721L..33L} {721, L33}

\bibitem[\protect\citeauthoryear{{Levin} et~al.,}{{Levin}
  et~al.}{2012}]{lbb+12}
{Levin} L.,  et~al., 2012, \mn@doi [MNRAS] {10.1111/j.1365-2966.2012.20807.x},
  \href {http://adsabs.harvard.edu/abs/2012MNRAS.422.2489L} {422, 2489}

\bibitem[\protect\citeauthoryear{{Liu} et~al.,}{{Liu} et~al.}{2015}]{lkl+15}
{Liu} K.,  et~al., 2015, \mn@doi [MNRAS] {10.1093/mnras/stv397}, \href
  {http://adsabs.harvard.edu/abs/2015MNRAS.449.1158L} {449, 1158}

\bibitem[\protect\citeauthoryear{{Liu} et~al.,}{{Liu} et~al.}{2016}]{lbj+16}
{Liu} K.,  et~al., 2016, \mn@doi [MNRAS] {10.1093/mnras/stw2223}, \href
  {http://adsabs.harvard.edu/abs/2016MNRAS.463.3239L} {463, 3239}

\bibitem[\protect\citeauthoryear{Lorimer \& Kramer}{Lorimer \&
  Kramer}{2005}]{lk05}
Lorimer D.~R.,  Kramer M.,  2005, Handbook of Pulsar Astronomy.
Cambridge University Press

\bibitem[\protect\citeauthoryear{{Lynch}, {Archibald}, {Kaspi}  \&
  {Scholz}}{{Lynch} et~al.}{2015}]{laks15}
{Lynch} R.~S.,  {Archibald} R.~F.,  {Kaspi} V.~M.,   {Scholz} P.,  2015,
  \mn@doi [ApJ] {10.1088/0004-637X/806/2/266}, \href
  {http://adsabs.harvard.edu/abs/2015ApJ...806..266L} {806, 266}

\bibitem[\protect\citeauthoryear{{Majid}, {Naudet}, {Lowe}  \&
  {Kuiper}}{{Majid} et~al.}{2011}]{mnl+11}
{Majid} W.~A.,  {Naudet} C.~J.,  {Lowe} S.~T.,   {Kuiper} T.~B.~H.,  2011,
  \mn@doi [ApJ] {10.1088/0004-637X/741/1/53}, \href
  {http://adsabs.harvard.edu/abs/2011ApJ...741...53M} {741, 53}

\bibitem[\protect\citeauthoryear{{Mori} et~al.,}{{Mori} et~al.}{2013}]{mgz+13}
{Mori} K.,  et~al., 2013, \mn@doi [ApJ] {10.1088/2041-8205/770/2/L23}, \href
  {http://adsabs.harvard.edu/abs/2013ApJ...770L..23M} {770, L23}

\bibitem[\protect\citeauthoryear{{Olausen} \& {Kaspi}}{{Olausen} \&
  {Kaspi}}{2014}]{ok14}
{Olausen} S.~A.,  {Kaspi} V.~M.,  2014, \mn@doi [ApJS]
  {10.1088/0067-0049/212/1/6}, \href
  {http://adsabs.harvard.edu/abs/2014ApJS..212....6O} {212, 6}

\bibitem[\protect\citeauthoryear{{Pennucci} et~al.,}{{Pennucci}
  et~al.}{2015}]{ppe+15}
{Pennucci} T.~T.,  et~al., 2015, \mn@doi [ApJ] {10.1088/0004-637X/808/1/81},
  \href {http://adsabs.harvard.edu/abs/2015ApJ...808...81P} {808, 81}

\bibitem[\protect\citeauthoryear{Rankin}{Rankin}{1986}]{ran86}
Rankin J.~M.,  1986, ApJ, 301, 901

\bibitem[\protect\citeauthoryear{{Rea} et~al.,}{{Rea} et~al.}{2010}]{ret+10}
{Rea} N.,  et~al., 2010, \mn@doi [Science] {10.1126/science.1196088}, \href
  {http://adsabs.harvard.edu/abs/2010Sci...330..944R} {330, 944}

\bibitem[\protect\citeauthoryear{{Rea} et~al.,}{{Rea} et~al.}{2012}]{rie+12}
{Rea} N.,  et~al., 2012, \mn@doi [ApJ] {10.1088/0004-637X/754/1/27}, \href
  {http://adsabs.harvard.edu/abs/2012ApJ...754...27R} {754, 27}

\bibitem[\protect\citeauthoryear{{Rea} et~al.,}{{Rea} et~al.}{2013}]{rep+13}
{Rea} N.,  et~al., 2013, \mn@doi [ApJ] {10.1088/2041-8205/775/2/L34}, \href
  {http://adsabs.harvard.edu/abs/2013ApJ...775L..34R} {775, L34}

\bibitem[\protect\citeauthoryear{{Rea}, {Vigan{\`o}}, {Israel}, {Pons}  \&
  {Torres}}{{Rea} et~al.}{2014}]{rvi+14}
{Rea} N.,  {Vigan{\`o}} D.,  {Israel} G.~L.,  {Pons} J.~A.,   {Torres} D.~F.,
  2014, \mn@doi [ApJ] {10.1088/2041-8205/781/1/L17}, \href
  {http://adsabs.harvard.edu/abs/2014ApJ...781L..17R} {781, L17}

\bibitem[\protect\citeauthoryear{Ritchings}{Ritchings}{1976}]{rit76}
Ritchings R.~T.,  1976, MNRAS, 176, 249

\bibitem[\protect\citeauthoryear{{Shannon} \& {Johnston}}{{Shannon} \&
  {Johnston}}{2013}]{sj13}
{Shannon} R.~M.,  {Johnston} S.,  2013, \mn@doi [MNRAS]
  {10.1093/mnrasl/slt088}, \href
  {http://adsabs.harvard.edu/abs/2013MNRAS.435L..29S} {435, L29}

\bibitem[\protect\citeauthoryear{{Spitler} et~al.,}{{Spitler}
  et~al.}{2014}]{sle+14}
{Spitler} L.~G.,  et~al., 2014, \mn@doi [ApJ] {10.1088/2041-8205/780/1/L3},
  \href {http://adsabs.harvard.edu/abs/2014ApJ...780L...3S} {780, L3}

\bibitem[\protect\citeauthoryear{{Thompson} \& {Duncan}}{{Thompson} \&
  {Duncan}}{1995}]{td95}
{Thompson} C.,  {Duncan} R.~C.,  1995, MNRAS, 275, 255

\bibitem[\protect\citeauthoryear{Thompson \& Duncan}{Thompson \&
  Duncan}{1996}]{td96a}
Thompson C.,  Duncan R.~C.,  1996, ApJ, 473, 322

\bibitem[\protect\citeauthoryear{{Tong}, {Yuan}  \& {Liu}}{{Tong}
  et~al.}{2013}]{tyl13}
{Tong} H.,  {Yuan} J.-P.,   {Liu} Z.-Y.,  2013, \mn@doi [RAA]
  {10.1088/1674-4527/13/7/007}, \href
  {http://adsabs.harvard.edu/abs/2013RAA....13..835T} {13, 835}

\bibitem[\protect\citeauthoryear{{Torne} et~al.,}{{Torne}
  et~al.}{2015}]{tek+15}
{Torne} P.,  et~al., 2015, \mn@doi [MNRAS] {10.1093/mnrasl/slv063}, \href
  {http://adsabs.harvard.edu/abs/2015MNRAS.451L..50T} {451, L50}

\bibitem[\protect\citeauthoryear{{Torne} et~al.,}{{Torne}
  et~al.}{2017}]{tde+17}
{Torne} P.,  et~al., 2017, \mn@doi [MNRAS] {10.1093/mnras/stw2757}, \href
  {http://adsabs.harvard.edu/abs/2017MNRAS.465..242T} {465, 242}

\bibitem[\protect\citeauthoryear{{van Straten} \& {Bailes}}{{van Straten} \&
  {Bailes}}{2011}]{vb11}
{van Straten} W.,  {Bailes} M.,  2011, \mn@doi [PASA] {10.1071/AS10021}, \href
  {http://adsabs.harvard.edu/abs/2011PASA...28....1V} {28, 1}

\bibitem[\protect\citeauthoryear{{Vivekanand}}{{Vivekanand}}{1995}]{viv95}
{Vivekanand} M.,  1995, MNRAS, 274, 785

\bibitem[\protect\citeauthoryear{{Wang}, {Manchester}  \& {Johnston}}{{Wang}
  et~al.}{2007}]{wmj07}
{Wang} N.,  {Manchester} R.~N.,   {Johnston} S.,  2007, \mn@doi [MNRAS]
  {10.1111/j.1365-2966.2007.11703.x}, 377, 1383

\bibitem[\protect\citeauthoryear{{Weltevrede}}{{Weltevrede}}{2016}]{wel16}
{Weltevrede} P.,  2016, \mn@doi [A\&A] {10.1051/0004-6361/201527950}, \href
  {http://adsabs.harvard.edu/abs/2016A%26A...590A.109W} {590, A109}

\bibitem[\protect\citeauthoryear{{Weltevrede}, {Edwards}  \&
  {Stappers}}{{Weltevrede} et~al.}{2006a}]{wes06}
{Weltevrede} P.,  {Edwards} R.~T.,   {Stappers} B.~W.,  2006a, A\&A, 445, 243

\bibitem[\protect\citeauthoryear{{Weltevrede}, {Wright}, {Stappers}  \&
  {Rankin}}{{Weltevrede} et~al.}{2006b}]{wws+06}
{Weltevrede} P.,  {Wright} G.~A.~E.,  {Stappers} B.~W.,   {Rankin} J.~M.,
  2006b, \mn@doi [A\&A] {10.1051/0004-6361:20065572}, \href
  {http://adsabs.harvard.edu/abs/2006A%26A...458..269W} {458, 269}

\bibitem[\protect\citeauthoryear{{Yan} et~al.,}{{Yan} et~al.}{2015}]{ysw+15}
{Yan} Z.,  et~al., 2015, \mn@doi [ApJ] {10.1088/0004-637X/814/1/5}, \href
  {http://adsabs.harvard.edu/abs/2015ApJ...814....5Y} {814, 5}

\bibitem[\protect\citeauthoryear{{Yao}, {Manchester}  \& {Wang}}{{Yao}
  et~al.}{2017}]{ymw17}
{Yao} J.~M.,  {Manchester} R.~N.,   {Wang} N.,  2017, \mn@doi [ApJ]
  {10.3847/1538-4357/835/1/29}, \href
  {http://adsabs.harvard.edu/abs/2017ApJ...835...29Y} {835, 29}

\makeatother
\end{thebibliography}


\bsp	
\label{lastpage}
\end{document}